\documentclass{article} 
\def \inbar{\vrule height1.5ex width.4pt depth0pt} 
\def \C{\relax\hbox{\kern.25em$\inbar\kern-.3em{\rm C}$}} 
\def \R{\relax{\rm I\kern-.18em R}}

\newcommand{\sgn}{{\rm sgn}}

 \newcommand{\beq}{\begin{equation}} 
\newcommand{\eeq}{\end{equation}}
 \newcommand{\bea}{\begin{eqnarray}}
 \newcommand{\eea}{\end{eqnarray}} 
\newcommand{\nn}{\nonumber}

\newcommand{\pdr}{\partial}

\newcommand{\Tr}{\hbox{Tr}}

\begin{document} 
\author{Erdal Toprak${\,}^{1}$ and  O. Teoman Turgut${\,}^{1,2}$\\  \\ ${}^{1}\,
$Department of Physics, Bogazici University \\ 
80815 Bebek, Istanbul, Turkey\\  
and\\ ${ }^{2}$Feza Gursey Institute\\ 
Kandilli 81220, Istanbul, Turkey\\   turgutte@boun.edu.tr}   
\title{\bf  Large N Limit of SO(N)  Scalar Gauge Theory } 
\maketitle 
\large 
\begin{abstract} 
\large  
In this paper we study the large $N_c$ limit of $SO(N_c)$ gauge theory coupled to 
a real scalar field following ideas of Rajeev\cite{2dqhd}.
We will see that the phase space of this  resulting classical theory is  
$Sp_1({\cal H})/U({\cal H}_+)$ which is the analog of the Siegel disc in infinite dimensions.
The linearized equations of motion give us a version of the well-known 
't Hooft equation of two dimensional QCD.

\end{abstract} 
\large 
\section{Introduction}  
Gauge theories are the fundamental theories that describe nature: Quantum
Chromodynamics 
 (QCD) is the gauge theory of hadrons, and it is believed that
 one can compute the 
 masses and excitations of all these hadrons from QCD.
As yet there is no satisfactory 
 understanding of these bound states. All the
hadrons are color singlets, in fact we never see
 the underlying quarks as
asymptotic states. That means QCD should be a confining theory and
 there
should be an independent formulation of it which is expressed completely in
terms 
 of these color invariant states.
 
In \cite{2dqhd} Rajeev has constructed such a theory of mesons in two dimensions in
the limit $N_c$, the number of colors in $SU(N_c)$, goes  to infinity. The idea that 
QCD should  simplify while keeping all its essential features  in this limit 
goes back to `t Hooft\cite{thooft1, thooft2} and that this limit 
should be a kind of classical 
mechanics to Migdal and Witten.
Even this large-$N_c$ theory is quite complicated and `t Hooft looked at a two dimensional 
model to understand the meson spectrum in this limit and obtained his bound state 
equation\cite{thooft2}. It is  not so clear how to treat the baryons in the large-$N_c$ limit. 
Witten suggested that baryons could also be understood 
(as solitonic excitations) in his by now classic papers\cite{witten, witten2}.
A much more ambitious program was presented by Lee and Rajeev\cite{leerajeev}
 for the large-$N_c$ limit 
of more complicated gauge theories.

In this paper we study the large $N_c$ limit of $SO(N_c)$ gauge theory coupled to 
a real scalar field. This theory is not physical, but it is a good model to test 
some of the ideas about gauge theories. We will apply the methods  developed  by Rajeev
to this toy model. We recommend the lectures of Rajeev for a more detailed 
exposition of the underlying ideas \cite{istlect}.
Rajeev in his work \cite{2dqhd} has shown that the phase space of the two dimensional QCD is an 
infinite dimensional Grassmannian, well-known from the theory of integrable systems and loop 
groups \cite{pressegal}. Scalar QCD was worked out using  the same methods in 
\cite{rajtur}, where it was shown that the 
phase space of the theory is an infinite dimensional disc. 
Originally scalar two dimensional QCD was worked out by Shei and Tsao 
in \cite{shei} following `t Hooft, and later by Tomaras using Hamiltonian
methods in \cite{tomaras}.
These works obtained the analog of the `t Hooft equation for this case.
A natural extension of these would be to look at combined (fermionic) QCD and scalar QCD,
 this is done 
in a paper of Aoki\cite{aoki} where it is shown that the various types of 
mesons are possible and they all obey  
`t Hooft equations. About the same time following a path integral approach and 
bilocal fields,  coupled fermions and bosons as well as some other  models 
in two dimensions are 
worked out in a beautiful article by
Cavicchi \cite{cavicchi}.
Recently Konechny and the second author have extended the methods 
of
\cite{2dqhd} to 
 the above case and showed that the underlying large-$N_c$
phase space is 
 a certain kind of super-Grassmannian.
The linearized equations agree with the ones found in \cite{aoki}. The correct 
equations are nonlinear, and there is a baryon number operator which 
corresponds to the 
supertrace of the basic variable\cite{tolyateo}.

The real scalar field is an interesting testing ground. There are some 
ideas in the literature which suggest that gauge theories in two dimensions all behave in a 
very similar way\cite{kutasov}. 
In this work we  show that the phase space of the   resulting classical theory is  
$Sp_1({\cal H})/U({\cal H}_+)$ which is the analog of the Siegel disc in infinite dimensions.
The linearized equations of motion give us a version  of the well-known 
't Hooft equation of two dimensional QCD, and this new one is the same equation found
in \cite{shei} apart from some numerical factors. That means that we have the same 
spectral behaviour for the mesons of the theory.
Since most of the details are very similar to the ones in Rajeev's lectures\cite{istlect} and 
various aspects of the geometry of the phase space was given  in a few  other  places
\cite{gracia, bowick,  segal} our tretment  will be brief.

\section{The scalar $SO(N_c)$ model in the light cone}
  
Since the basic philosophy was explained in \cite{2dqhd} we will state our conventions and 
write down the Lagrangian of the theory.
 We will use the light cone coordinates  
$x^{+} = \frac{1}{\sqrt{2}}( t+x)$,  $x^{-} = \frac{1}{\sqrt{2}}( t-x)$  and 
choose   $A_{-} = 0$ gauge.   
\beq
S =  \int dx^{+}dx^{-}\bigl[ {1 \over 2}\Tr F_{+-}F^{+-} + 
{1\over 2} \phi^{\alpha}(-2\pdr_-)\pdr_+\phi_{\alpha}+ 
{g\over 2} (\pdr_-\phi^{\alpha} {A_+}^\beta_\alpha
\phi_\beta-\phi^{\alpha}{A_+}^\beta_\alpha
\pdr_-\phi_\beta)- {m^2\over 2}\phi^{\alpha}\phi_\alpha \bigr]  
.\eeq 
Here we have $SO(N_c)$ gauge theory for which  the matter fields 
are in the fundamental representation and Tr denotes an invariant inner product in the 
Lie algebra.  
The Lie algebra condition implies that $A_+^T=-A_+$.
To compute the variations we need the independent degress of freedom,
we can expand $A_+=A_+^aT^a$ where $T^a$ are the generators of $SO(N_c)$ Lie algebra.
We can choose them such that Tr$T^a T^b=-{1\over 2}\delta^{ab}$. 
When we use the light cone approach in $1+1$ dimensions, 
 the gauge fields  do not carry dynamical degrees of freedom.
We  first eliminate the gauge fields and then write the resulting action.
Let us find the  equation of motion  for the gauge field once we rewrite  the action.

\beq 
S =  \int dx^{+}dx^{-}\bigl[ {1\over 2} \phi^{\alpha}(-2\pdr_-)\pdr_+
\phi_{\alpha}+ 
{g\over 2} A_+^a(\pdr_-\phi^{\alpha} {T^a}^\beta_\alpha
\phi_\beta-\phi^{\alpha}{T^a}^\beta_\alpha
\pdr_-\phi_\beta)  +{1\over 2} (\pdr_-A_+^a)^2- {m^2\over 2}
\phi^{\alpha}\phi_\alpha \bigr]  
.\eeq  
If we define the current 
$J^a ={1\over 2} (\phi^\alpha {T^a}^\beta_\alpha \pdr_-\phi_\beta-
\pdr_-\phi^\alpha {T^a}^\beta_\alpha\phi_\beta)$,
we get 
\beq
      -\pdr_-^2A_+^a=g J^a
,\eeq  
which can be solved  formally (an actual solution can be found if we specify 
some  boundary conditions) and by substituting this into our Lagrangian again,
\beq
    S=\int dx^+dx^- \Big( {1\over 2} \phi^{\alpha}(-2\pdr_-)\pdr_+\phi_{\alpha}
+[{g^2\over 2} J^a {1\over \pdr_-^2} J^a + {m^2 \over 2} \phi^\alpha
\phi_\alpha]\Big)
 .\eeq
Written in  this form we immediately see that \cite{heinzl, 2dqhd} we have the 
following  symplectic form 
\beq
     \omega^{-1} (x^-,y^-)=< x^-| (-2\pdr_-)^{-1} | y^->=-{1 \over 4} {\rm sgn}(x^--y^-)
\eeq
and the Hamiltonian,
\beq
     H=\int dx^- \Big[-{g^2\over 2} J^a {1\over \pdr_-^2} J^a + 
{m^2 \over 2} \phi^\alpha \phi_\alpha\Big]
.\eeq
The same boundary conditions as the one used to find the 
symplectic form gives us the more explicit expression,
\beq
  H={m^2 \over 2} \int dx^-\phi^\alpha(x^-)\phi_\alpha(x^-)
 -{g^2\over 4} \int dx^-dy^- J^a(x^-) |x^--y^-|J^a(y^-)
\eeq
We note that one  needs the properties of the group and its 
represention to compute the interaction term.
This can be achieved due to
the identity 
$\sum_a (T^a)_{\alpha\beta} (T^a)_{\lambda\gamma}= -{1\over 2} 
(\delta_{\alpha\gamma} \delta_{\lambda\beta}-\delta_{\alpha\lambda}\delta_{\beta\gamma})$.

It is now possible to   compute the equations of motion for the classical variable
$\phi(x^-;x^+)$ using 
\beq
        {\partial \phi\over \pdr x^+} =\{ \phi, H\}
.\eeq
It is a useful exercise to find  
the equations of motion even for the free field theory( see the beautiful lectures by 
Heinzl \cite{heinzl}) . Another important exercise is to check that the theory is Poincare 
invariant written in this new way, by finding the generators.

We will follow \cite{2dqhd} (or \cite{istlect}) and rewrite the
 theory in terms of the color invariant 
bilinears of the 
field variable $\phi$ after canonical quantization.
In the large $N_c$ limit  these will be the only dynamical variables,
and the theory has a completely classical formulation in terms of these 
bilinears.
We will see that the remaining global $SO(N_c)$ symmetry we have 
imposes a constraint for these variables and that means the phase space of the 
theory is a curved manifold in infinite dimensions.

Canonical quantization is standard, since the theory is super-renormalizable 
the result is the same as free field theory and the choice of vacuum is 
exactly the same. The equal ``time" commutator is given by
\beq
       [\hat \phi_\alpha(x^-,x^+), \hat \phi_\beta(y^-,x^+)]=-{i\over 4} \delta_{\alpha\beta}
{\rm sgn}(x^--y^-)
.\eeq
This means that it is simpler to introduce creation and annihilation operators,
satisfying,
\beq
     [a_\alpha(p),a_\beta(q)]= 2\pi \delta(p+q)\delta_{\alpha\beta}\sgn(p)
\eeq
such that 
\beq
    \hat \phi_\alpha(x^-)=\int {dp\over 2\pi} {a_\alpha(p) \over
\sqrt{2|p|}}e^{-ipx^-}
 .\eeq
For quantum  theory we introduce the Fock vacuum $|0>$:
\beq
         a_\alpha(p)|0>=0 \qquad {\rm when}\  0\leq p
\eeq 
To get well-defined expressions for various operators--
such as the Hamiltonian--
we need a normal ordering prescription:
\beq
      :a_\alpha(p)a_\beta(q):=\cases{a_\beta(q)a_\alpha(p) \  {\rm if \ } q<0, p>0 \cr 
                             a_\alpha(p)a_\beta(q)  \ {\rm otherwise} }
\eeq 
We note that it is also possible to  think about   the creation and
annihilation operators via a Fourier expansion,
\beq
  \hat \phi_\alpha(x^-)=\int_0^\infty {dp \over 2\pi
\sqrt{2|p|}} (a_\alpha(p) e^{ -ipx^-}+ a_\alpha^\dag(p)e^{ipx^-})  
,\eeq
manifesting the real valuedness of the field $\hat \phi^\dag_\alpha=\hat \phi_\alpha$.
This automatically implies that 
$a^\dag_\alpha(p)=a_\alpha(-p)$ for $p>0$.
In the next section we will see that this is a more appropriate way to think about 
quantization, yet from a calculational point of view the other is better. 
One notes that this is consistent with the commutation relations of $a_\alpha$'s.
(See \cite{heinzl} and \cite{brodsky} for more details about the light-cone vacuum structure of 
the real
scalar field).

The normal ordering can be written in terms of the ordinary products of the operators and 
a vacuum subtraction,
\beq
   :a_\alpha(p)a_\beta(q): = a_\alpha(p) a_\beta(q) -{1\over 2}({\rm sgn}(p)+1)2\pi 
\delta(p+q)\delta_{\alpha\beta}
.\eeq
We will make use of this relation quite frequently in the next section.

\section{ Algebra of Color Invariant Operators}

In this section we will discuss the class of operators we will use to 
reformulate the gauge theory in the large-$N_c$ limit.
Since we have fixed the gauge as $A_-=0$ we are not allowed to make any more 
space dependent gauge transformations.
(The equations of motion at the quantum level imply that the ``time" dependent
transformations cannot be made arbitraily but given by the evolution of 
the scalar field. We do not need to look at these in any case since in the Hamiltonian 
formalism observables at a fixed ``time" slice are enough).
Yet there is still a global $SO(N_c)$ symmetry which is left over.
To emphasize the contraction we write down the color invariant bilinears
with one index up the other index down,
\beq 
        N(x^-,y^-)={1\over N_c} :\hat \phi^\alpha(x^-)\hat \phi_\alpha(y^-):
.\eeq
The set of these equal time bilinears constitute the set of all possible 
color invariant operators for this theory.
One may equally look at the Fourier transform of these operators,
so the basic bilinears in this case become,
\beq
    \hat  T(p,q)={2\over N_c}\sum_\alpha :a_\alpha(p)a_\alpha(q):
\eeq
As we will see in the next section that conceptually it is more natural to use the 
variables 
\beq
   \hat K(p,q)=-{2\over N_c} \sgn(p) \sum_\alpha :a_\alpha^\dag(p) 
a_\alpha (q):\quad ,
\eeq
but for calculations it is easier to keep the above variables. 
The basic idea of the large-$N_c$ theory is to write everything in terms of these color 
invariant bilinears. In the limit $N_c$ becomes large 
only the color invariant operators survive and 
furthermore the expectation values of color invariant operators split as a porduct upto 
$1/N_c$ corrections. This implies that the set  of color invariant operators 
becomes  classical, all color invariant operators should be 
representable as classical  observables.
The resulting theory, restricted to the space of color invariant states
therefore becomes a classical theory \cite{2dqhd, istlect, yaffe}.
To define this classical theory we compute the commutator of two 
such color invariant operators and then take the appropriate large-$N_c$ limit.
The result will be postulated as a classical Poisson 
bracket of these classical 
variables.
We will see later on that this Poisson bracket actually comes from a symplectic form 
on a very natural infinite dimensional homogeneous symplectic manifold \cite{bowick}.

When we compute the commutator
of such bilinears we get,
\bea
&\ & [\hat T(p,q), \hat T(s,t)]= \nn\cr
&\ & \ {2\over N_c}\Big( {\rm sgn} (p) \delta[p+s]\hat  T(q,t)
+{\rm sgn}(q)\delta[q+s] \hat T(p,t)+{\rm sgn}(p)\delta[p+t]\hat T(s,q)+{\rm sgn}(q)\delta[q+t]
\hat T(s,p)\nn\cr
 &\ & \qquad \quad + ({\rm sgn} (p)+{\rm sgn} (q) )(\delta[p+s]\delta[q+t] 
+\delta[p+t]\delta[s+q])\Big),
\eea
where we defined $\delta[p+q]=2\pi \delta(p+q)$ for convenience.

If we take the limit $N_c\to \infty$
we assume that there are corresponding classical observables  and 
the commutators go to Poisson brackets of these observables.
We still denote them by the same letter except  dropping the hat on the top.
Applying the rule $-{i\over \hbar} [A, B]\mapsto \{ A, B\}$,
as $\hbar=N_c^{-1}\to 0$, we get\footnote{There is really no way to determine the correct 
quantization parameter in this approach. We can only find this when we quantize the theory
back again. The most natural method to employ is geometric quantization, due to the natural 
geometry of the phase space. We will come back to these issues in a separate publication.},

\bea
&\ & \{ T(p,q),  T(s,t)\}= \nn\cr
&\ & \ -2i\Big( {\rm sgn} (p) \delta[p+s]  T(q,t)
+{\rm sgn}(q)\delta[q+s] T(p,t)+{\rm sgn}(p)\delta[p+t] T(s,q)+{\rm sgn}(q)\delta[q+t]
 T(s,p)\nn\cr
 &\ & \qquad \quad + ({\rm sgn} (p)+{\rm sgn} (q) )(\delta[p+s]\delta[q+t] 
+\delta[p+t]\delta[s+q])\Big).
\eea
We will postulate these to be the basic Poisson brackets of our dynamical 
variables. 
It is a good exercise to compute the equations of motion for the free field in this 
language and write down the solution.

These variables acting on the color invariant sector are not completely independent,
there is a constraint coming from the global color invariance.
Recall that the global $SO(N_c)$ is generated by the operators acting on the
Fock space,
\beq
   \hat Q_{\alpha\beta} =\int_0^\infty {dp\over 2\pi} a^\dag_\alpha(p)a_\beta(p)
-\int_0^\infty {dp \over 2\pi} a_\beta^\dag (p) a_\alpha(p)
.\eeq
These operators satisfy $\hat Q_{\alpha\beta}|0>=0$ and 
\beq
   [\hat Q_{\alpha\beta} , \hat Q_{\lambda\gamma}]=
\hat Q_{\lambda\alpha}\delta_{\beta\gamma} +\hat Q_{\beta\lambda} \delta_{\alpha\gamma}
+\hat Q_{\gamma\beta}\delta_{\lambda\alpha} +\hat Q_{\lambda\alpha}\delta_{\gamma\beta}
.\eeq
One can see that  $\hat Q_{\alpha\beta}=-\hat Q_{\beta\alpha}$.
Recall the related  set of bilinear  variables, 
\beq
\hat K(p,q)=-{ 2\over N_c} \sgn(p): a^\dag_\alpha(p)a_\alpha(q):,
\eeq
a careful computation shows that when we restrict 
these variables to the color 
invariant sector of the Fock space in the large-$N_c$ limit we get,
\beq
\int_{-\infty}^{\infty}  K(p,s) K(s,q)[ds]-\sgn(p) K(p,q)-
 K(p,q)\sgn(q)=0.
\eeq
This operator equation is interpreted as now anequation for the 
kernel of an integral operator acting on the 
one-particle Hilbert space.
We can write the same constraint in a more succint manner as,
\beq
       ( K+\epsilon)^2=I
,\eeq
where $\epsilon(p,q)=-\sgn(p)\delta[p-q]$ and we interpret this as an operator
equation again. We will talk about the meaning of this equation from 
a more geometric point of view  in the next section. 
The important assumption is that when we let $N_c\mapsto \infty$,
the above constraint translated into a constraint for the 
classical variables $K$. So the dynamical variables $K$
satisfy this constraint, which  implies a constraint for  $T(p,q)$
trivially.

We rewrite the Hamiltonian by redefining the coupling constant as 
$g^2N_c\mapsto g^2$ and dividing the Hamiltonian 
by an overall factor of $N_c$. 
Thus the Hamiltonian becomes, after mass renormalization,
\beq
    H_0={1\over 8} (m_R^2-{g^2\over 2\pi}){\cal P}\int {[dp]\over |p|} T(-p,p)
,\eeq
where the renormalized mass is
given by $m^2=m^2_R+{g^2\over 4\pi}\ln {\Lambda_U\over\Lambda_I}$, $\Lambda_U,\Lambda_I$ refering to the 
ultraviolet and infrared cut-offs respectively, 
 also 
 we used the shorthand $[dp]={dp\over 2\pi}$, and 
${\cal P}$ denotes the principal value prescription.
This is not a simple computation but the 
essential steps are given in \cite{istlect}, 
 and the interaction part
\beq
    H_I= {g^2\over 64}{\cal FP}\int {[dpdqdsdt]\over \sqrt{|pqst|}}\delta[p+q+s+t]
{sq-st+pt-pq\over (p+s)^2}T(p,q)T(s,t)
,\eeq
where ${\cal FP}$ denotes the finite part, as explained in 
\cite{istlect}. For simplicity of notation from now on 
we will drop the symbols, ${\cal P}$ and ${\cal FP}$, but 
the calculations should be performed keeping these in mind.
At this point we have the complete formulation of our theory, one can  compute the
equations of motion using the above form of the Hamiltonian and the Poisson brackets 
of the variables $T(u,v)$. At this stage we will not be able to give an analysis of these 
nonlinear equations and instead confine ourselves to the linear approximation.

For the linear approximation we follow 
\cite{istlect}, we will write the  above  constraint in terms of the 
$T$ variables and ignore the second order term in $T$. This will be our linearized constraint,
\beq
    [1-{\rm sgn} (u){\rm sgn} (v)]T(u,v)=0
.\eeq
In the following we will keep all the equations of motion to this approximation and 
search for a bound state solution.

We can compute the equations of motion in the linear approximation:
this means we look at $T(u,v)$ for $u,v>0$ or $u,v<0$, the other cases imply 
$T(u,v)=0$ from the constraint equation.
Let us look at $u,v<0$ case and define 
$P=-(u+v)$ and $x=-u/P$.
This means $u=-Px,v=-P(1-x)$ and $0<x<1$.
If we actually compute the equations of motion 
$\partial_+T(u,v;x^+)=\{T(u,v;x^+), H\}$, and make an ansatz,
$T(u,v)=e^{iP_+x^+}\zeta(x)$,
we get 
\bea
   &\ &   \mu^2\zeta(x)=(m_R^2-{g^2\over 2\pi})[{1\over x} +{1\over 1-x}]\zeta(x)\nn\cr
   &\ &-{g^2\over 8\pi}\int_0^1 \Big[{y(1-x)+x(1-y)+y(1-y)+x(1-x)\over (x-y)^2}\nn\cr
   &\ &  +{xy+(1-x)(1-y)+y(1-y)+x(1-x)\over (y-(1-x))^2}\Big]
 { \zeta(y)dy \over \sqrt{x(1-x)y(1-y)}}
,\eea
where $\mu^2=2P_+P$ is the invariant mass of this excitation.
We should solve this eigenvalue equation to find the allowed values of $\mu^2$ and 
the function $\zeta$. This will determine the spectrum of the theory.  
One notes that the equation is symmetric under $x\mapsto 1-x$ and $y\mapsto 1-y$, that 
means we may choose $\zeta(x)=\zeta(1-x)$.
This simplifies  our equation to 
\beq
      \mu^2\zeta(x)=(m_R^2-{g^2\over 2\pi})[{1\over x} +{1\over 1-x}]\zeta(x)
-{g^2\over 4\pi}\int_0^1 {(x+y)(2-x-y)\over (x-y)^2}
{ \zeta(y)dy \over \sqrt{x(1-x)y(1-y)}}
.\eeq
The above form is in fact identical to the bound state equation found in reference 
\cite{shei} and later on by Tomaras using Hamiltonian methods apart from the 
numerical factors (this approach is closer 
 to the one in \cite{2dqhd}).
It is known that this theory has only discrete states, that is we only have bound state 
solutions and no scattering states. 

We may search for the  baryons in this theory (from a more standard point of view,
we do not have any $U(1)$ symmetry in the classical action, this suggests 
that there should not be baryon number conservation and no baryons, we will
see that the baryon number is indeed not conserved for the gauge theory).
 
Note that there is no anti-baryon. Let us write 
 down a typical baryonic
operator;
 \beq
    B(p_1,p_2,...,p_{N_c})= {1 \over Z} \epsilon_{\alpha_1\alpha_2...\alpha_{N_c}}
a^\dag_{\alpha_1}(p_1) a^\dag_{\alpha_2}(p_2)...a^\dag_{\alpha_{N_c}}(p_{N_c})
,\eeq
where $Z$ is an appropriate normalization factor. When we take the large-$N_c$ limit
these operators become infinite strings which are not representable in a simple way.
But we can still detect them if they are present in a physical state.
We write a one-baryon state as 
$B(p_1,p_2,...,p_{N_c})|0>$, and define  the baryon operator,
\beq
\hat B={1\over N_c}\int_0^\infty [dp] :a^\dag_\alpha(p) a_\alpha(p):
.\eeq
In general we have the action of the baryon operator on many baryon 
states,
\bea
      {1\over N_c}\int_0^\infty &[dp]& :a^\dag_\alpha(p)a_\alpha(p):
B(p_1,p_2,...,p_{N_c})B(q_1,q_2,...,q_{N_c})...B(s_1,s_2,...,s_{N_c})|0>=\nn \cr
&\ &{\rm  \Big({\bf number\  of\  baryons\ }\Big)}B(p_1,p_2,...,p_{N_c})
B(q_1,q_2,...,q_{N_c})...B(s_1,s_2,...,s_{N_c})|0>
.\eea
We may have mesonic parts in general, but in this 
picture they seem to  be of smaller order.
Note that this operator will survive the large-$N_c$ limit and 
can be represented as the half trace of the variable $T(p,q)$ 
evaluated only for 
the positive momenta.
A natural question is if  the baryon number operator is conserved
under the evolution of our system--
it does not follow from a simple symmetry principle--A
direct method is to see if this operator Poisson commutes with a
quadratic Hamiltonian.
Let us write down a general quadratic Hamiltonian as
\beq
   H=\int [dp] h(p) T(-p,p)+\int [dpdqdsdt] G(p,q;s,t)T(p,q)T(s,t)
.\eeq
The Poincare invariance will impose 
certain restrictions on the choice of functions $h,G$. There are 
   a few obvious 
symmetries  coming 
from the properties of the variable $T$, 
the considerations of the next section shows all
the symmetries required on
$G(p,q;s,t)$. If we compute now 
\beq
     \{ H, \int_{-\infty}^{\infty} T(-u,u)[du]\}=2i\int [\sgn(p)+\sgn(q)] G(p,q;s,t)
T(p,q)T(s,t)[dpdqdsdt]
,\eeq
the use of  the symmetries in general will not  
give zero:
this means that the baryon number is not conserved in general!
In our case the computation gives a nonzero result,
thus in the conventional sense {\it we do not have baryons}, yet we may have 
nonzero values of the trace implying possible baryonic states. 
We will see more comments on this from the geometry in the next section.

\section{Geometry of the Phase Space}

In this section we  present a somewhat more rigorous approach and provide an 
interpretation of the underlying phase space of the theory.
To do this let us discuss quantization again, for this we closely follow the 
ideas in the article by Bowick and Rajeev\cite{bowick} and for a more detailed 
presentation we refer to the beautiful  article  by 
Gracia-Bondia and Varilly\cite{gracia}. There is also a nice representation theoretic 
presentation in \cite{neeb}.

When we look at a real scalar field in two dimensions in the light cone formalism, we may 
formally quatize the system by declaring existence of operators corresponding to the
fields and we replace   
the Poisson bracket relations of these fields by commutators with an additional factor of 
$i$. Of course we assume that there is an underlying
{\it complex} Hilbert space, on which these operators act! In this formal process we 
do not see where the complex structure comes from. 
In fact there is a natural complex structure:
let us assume that the free hamiltonian is formally written 
as a quadratic form in the fields,
$H_0=\int {1\over 2} \phi_\alpha Q_{\alpha\beta} \phi_\beta$,
and we have a symplectic structure,
$\omega$, 
$\int {1\over 2} \phi_\alpha \omega_{\alpha\beta}\partial_+\phi_\beta$. 
This symplectic structure defines a skew form on the 
space of solutions to the classical field 
equations.
The natural operator to 
introduce  is  $\tilde \omega=\omega^{-1} Q$, this is a real  antisymmetric
operator(matrix) of type $(1,1)$, and comes from 
the equations of motion. We 
use its polar decomposition,
$\tilde \omega=J S$, where $J^TJ=1$ and $S^T=S$ with $S>0$.
Now using the antisymmetry of $\tilde \omega$ we  see that $J^2=-1$.
This defines a complex structure which we can use to complexify our real Hilbert space.
If we apply this to our case,
the metric coming from the free Hamiltonian, 
$H_0={m^2\over 2} \int dx^- \phi_\alpha(x^-)\phi_\alpha(x^-)$,
becomes,   $Q_{\alpha\beta}(x^-,y^-)=m^2\delta(x^--y^-)\delta_{\alpha\beta}$, 
and the symplectic form (see the previous section)
$\omega_{\alpha\beta}(x^-,y^-)=<x^-|-2\partial_-|y^->\delta_{\alpha\beta}$.
If we write down the polar decomposition, we have,
\beq
J_{\alpha\beta}(x^-,y^-)=<x^-|-(\partial_-^T\partial_-)^{1/2}
\partial_-^{-1}|y^->\delta_{\alpha\beta}
=<x^-|-(-\partial^2_-)^{1/2}\partial_-^{-1}|y^->\delta_{\alpha\beta}.
\eeq
Written in this form this is a real operator  acting on the 
$L_2$ space of initial data on the light cone.
We can extend this operator to a complex Hilbert space and it  is then possible to diagonalize
 the above $J$ in this  complexified space.
So we think of a complex $L_2$ space,
$V^{\bf C}=V\otimes {\bf C}=W\oplus\bar W$, where $W$ 
is isomorphic to $\bar W$, 
in the infinite dimensional case 
they are both separable.   The decomposition we use 
corresponds to the eigenspaces of $J$. If we write $J$ as 
a block diagonal on such a 
decomposition we get $J=\pmatrix{i&0\cr 0&-i}$.
We know from our experience in physics that this is the form we use.
If we decompose the fields into Fourier modes at the initial data surface $x^+=0$,
\beq
    \phi_\alpha(x^-)=\int_0^\infty {[dp]\over \sqrt{2 p}} \Big
(\bar z_\alpha(p)e^{-ipx^-} + z_\alpha(p)e^{ipx^-}\Big )
\eeq
and act upon it by $J$, we see that we get 
\beq
     (J\phi_\alpha)(x^-)=\int_0^\infty {[dp]\over \sqrt{2 p}} 
\Big(-i\bar z_\alpha(p)e^{-ipx^-}+i z_\alpha(p)e^{ipx^-}\Big)
.\eeq
So we see that the decomposition of the field into its positive and negative frequency modes 
is the same as using the eigenvalue decomposition of the underlying complex structure.
(We note that this decomposition is relativistically invariant, and the 
division by momentum variable $\sqrt{2p}$ is for convenience).
Now we can also see that the inverse of our skew form 
transforms under such a change of basis as
$R^{-1}\omega^{-1} (R^{-1})^T$, where we represent the Fourier transform 
as $R$,  here  $T$ refers to the ordinary transpose.
Thus we evaluate, 
\beq
\int dx^-dy^-e^{iqx^-}\sqrt{2|q|}({-1\over 2\partial_-})e^{ipy^-}\sqrt{
2|p|} = i\sgn(p)\delta[p+q]
,\eeq 
which shows that the 
symplectic form transforms to the standard form now defined on a 
complex Hilbert space.

The correct quantization in the infinite dimensional case requires this complex structure,
the formal quantization rule,
\beq
     [\hat \phi_\alpha(x^-), \hat \phi_\beta(y^-)]=-{i\over 4}\delta_{\alpha\beta}\sgn(x^--y^-)
,\eeq
clearly  requires a complex space, we assume  the real field is a self-adjoint
operator, $\hat \phi^\dagger(x^-)=\hat \phi(x^-)$.
In fact we really think of this system in terms of creation and 
annihilation operators acting on a complex Hilbert space.
This is best done by going into a Fourier decomposition and introducing 
the creation and annihilation operators corresponding to positive and 
negative frequency components.
Such a decomposition is necessary to make the commutation relations 
meaningful, a glance at them shows that 
$[a_\alpha(p), a_\beta(q)]=\sgn(p)\delta[p+q]$ is consistent 
with the creation and annihilation operator interpretation if 
we define $a_\alpha(p)$ to be the annihilation and 
$a_\alpha(-p)$ to be the creation operators for $p>0$. Now we see 
that what determines this
is precisely the complex structure, $J=-i\sgn(p)$.
This  form of the complex structure reveals another important aspect 
of this problem: there is no dependence on the mass. 
{\it If the bare  mass  changes due to the 
interactions, this does not change the  quasi-free representation 
of our commutation relations that were chosen  at the 
start  using the free part only}. 
The frequencies obviously 
change but that does not affect the representation.
To make the Hamiltonian and various 
other operators of physical interest well-defined in this Fock space 
we must introduce a normal ordering prescription.

If we compute the commutator of two normal ordered bilinears 
of the field operators, that provides a realization of the
real Symplectic Lie algebra in
its standard form.
When we switch to the Fourier modes, and use  the corresponding creation and 
annihilation operators we use the embedding of the
real symplectic Lie algebra into the complex symplectic Lie algebra.
In fact our operators $K(p,q)$, in the large-$N_c$ limit,  correspond to the 
Lie algebra generators with respect to this embedding, we will discuss this
below.
If we define our symplectic form as a matrix
$\omega=\pmatrix{0&1\cr -1&0}$, and the 
complex structure as $J=\pmatrix{0&1\cr -1&0}$,
we can diagonalize our complex structure in a complex Hilbert space
by $R={1\over \sqrt{2}}\pmatrix{i&-i\cr 1&1}$, then 
$R^{-1}JR=\pmatrix{-i&0\cr 0&i}$, whereas 
$R^T\omega R=i\pmatrix{0&1\cr -1&0}$.
In such an embedding the real symlectic group defined 
by $\omega$ becomes,
\beq
      \pmatrix{a&b\cr \bar b&\bar a}
,\eeq
and naturally still preserves the transfomred form of $\omega$,
but that is the same as the complex symplectic group, since 
$\omega$ as a matrix preserves its form.
A general complex symplectic matrix $g=\pmatrix{a&b\cr c&d}$,  satisfies,
\beq 
        a^T c=c^T a\quad b^Td=d^T b\quad a^Td-c^Tb=1
.\eeq
In our example we see that the Fourier transform does this
transformation: it brings 
$J$ into diagonal form and $\omega$ to the standard form.

The real Lie algebra can be written as
\beq
      1+i\pmatrix{F&G\cr -\bar G& -\bar F}
,\eeq
where $F^\dag =F$ and $ G^T=G$.
In fact one can check that the large-$N_c$ limit operators $K(p,q)$ obey these conditions.
Furthermore there will be convergence conditions coming from the super-renormalizability 
of our theory.
This corresponds to the fact that we require normal ordered bilinears to 
create finite norm states when they act on any other state constructed from the vacuum by the 
action of creation operators--of course strictly speaking we should think about
smeared out operators but we will ignore this technical part for this work. 
We can simply say that the off-diagonal components of these operators, that is
$b$ parts, should be Hilbert-Schmidt operators. In the same way we demand the same for the 
Lie algebra elements.( In higher dimensional theories this requirement is not satisfied and 
one needs a much more sophisticated not completely understood approach. One possibility was 
proposed by Mickelsson and Rajeev\cite{mickraj, mickbook}).
 
In this language the constraint should be written as 
$(iK+i\epsilon)^2=-1$, and $i\epsilon= J$, i.e. it is the diagonal form 
of the complex structure we were to begin with. There is theskew form 
which has a   matrix form in this basis 
$ \omega=\pmatrix{0&1\cr -1&0}$ which defines the symplectic group on
$W\oplus\bar W$.
We will see that the constraint actually 
defines a homogenenous manifold of the 
underlying real  symplectic group. 
If we introduce a variable $\Phi=K+\epsilon$, the constraint becomes,
\beq 
\Phi^2=1.
\eeq
One can also verify the following condition,
\beq
\Phi^T=\omega^{-1} \Phi \omega.
\eeq
This is nothig but the Lie algebra condition. In this 
basis there is no difference as matrices between 
$\omega$ and $\omega^{-1}$ but we should remember that 
they transform differently.
 Furthermore the convergence 
condition becomes,
\beq
[\epsilon, \Phi]\in  {\rm Hilbert-Schmidt}.
\eeq
As we will see in the following part these conditions correspond to the infinite dimensional 
version of the Siegel disc. 

We now   define a homogeneous manifold which will be denoted by 
$D_1^{R}$. It is essentially a real version of the disc which corresponds to the
pseudo-unitary group.
Let us define a Hilbert space ${\cal H}_+$,  which refers to the positive
frequency modes of the
theory. We can also say that these are the functions which have 
only positive modes in their Fourier decomposition.  We introduce a set of 
operators 
$Z:{\cal H}_+ \to {\cal H}_-$, where ${\cal H }_-$ is $\bar{{\cal H}_+}$ in the above language.
(If we  use the full complex Hilbert space, ${\cal H}={\cal H}_+\oplus{\cal H}_-$). We impose
$Z^T=Z$. We have a complex conjugation $\sigma$, this 
intertwines between ${\cal H}_+$ and ${\cal H}_-$, we define 
$Z^T=\sigma Z^\dag \sigma$, note that $Z^\dag:{\cal H}_-\to 
{\cal H}_+$, thus $Z^T:{\cal H}_+\to {\cal H}_-$. Furhermore 
$\bar Z=\sigma Z\sigma:{\cal H}_+\to {\cal H}_-$. There is an
extra  condition on $Z$:
$1-Z^\dag Z>0$. We also need a convergence condition which 
comes from the infinite
dimensionality of the theory:
$Z \in {\cal I}_2$, where ${\cal I}_2$ denotes the Hilbert-Schmidt ideal \cite{simon, istlect,
mickbook}.

We introduce a real restricted  symplectic group, $Sp_1$ embedded into the
above mentioned complex symplectic group, which we precisely define  below:
\beq
    Sp_1^c({\cal H})=\{ g:{\cal H}\to {\cal H}| g^{-1} \ {\rm exits,}\
    g^T \omega  g=\omega\ 
{\rm and }\ [\epsilon, g]\in {\cal I}_2 \}
,\eeq
here we are using ordinary matrix transpose to be able to write 
explicit matrix  elements.
We can see that this is a group and we call it the restricted complex
symplectic group, and its subgroup of the form 
\beq
     \pmatrix{a &b\cr \bar b& \bar a} \quad b\in {\cal I}_2
\eeq
corresponds to  the restricted real symplectic group 
$Sp_1({\cal H})$. $J$ itself is a real 
symplectic matrix and we are using a basis 
for the complexified Hilbert space in which 
$J$ becomes diagonal.

The real  symplectic group has an action on the space of operators 
$Z$,  given by 
\beq
    g\circ Z=(aZ+b) (\bar b Z+\bar a)^{-1}
.\eeq
We can check that the action obeys the usual rule
$g_1\circ (g_2\circ  Z)=(g_1g_2)\circ Z$.
To prove that the action preserves all the conditions we look at the 
orbit of $Z=0$, which is obviously in this set $D_1^R$.
We see that the the resulting element 
$b\bar a^{-1}$ satisfies all the properties, hence the orbit remains inside the disc.
(We should of course see that the inverse of $\bar a$ exists, but that is easy using the 
properties of the group).
Let us assume that a $Z$ is given, we claim that any such element lies in the orbit of
$Z=0$. To show this we explicitly construct a group element which does this:
\beq
    g(Z)=\pmatrix{ (1-\bar Z Z)^{-1/2} & Z(1-\bar Z Z)^{-1/2}\cr
        \bar Z(1-Z\bar Z)^{-1/2} & (1-\bar Z Z)^{-1/2}}
,\eeq
note that everything here is well-defined.
We leave it to the reader to check that $g(Z)$ is an element of the 
real group. 
This shows that the disc is actually a homogeneous space: take any element 
$Z$, pull it back to $Z=0$, by $g^{-1}(Z)$ and to reach any element 
$\tilde Z$ use the group element corresponding to this for the orbit of $Z=0$,
$g(\tilde Z)$ and use the compatibility condition,
$\tilde Z= (g(\tilde Z)g^{-1}(Z))\circ Z$. It is clear that the action 
then remains inside the disc.

We see that the disc is actually a complex homogeneous space, the stability subgroup 
corresponding to $Z=0$ is given by 
\beq
     U({\cal H}_+)=\pmatrix{a &0\cr 0 &\bar a}
.\eeq
If we use  symplectic condition  we get $a^\dag a=a a^\dag=1$, which means 
$a$ is an element of the unitary group of ${\cal H}_+$.
This means we have 
\beq  
   D^R_1={ Sp_1({\cal H})\over U({\cal H}_+)}
.\eeq

We will in fact see that the above space is a complex homogeneous symplectic 
manifold, but before this it is useful to introduce a variable 
$ \Phi(Z)$:
\beq
\Phi(Z)= -1 +2\pmatrix{(1-Z\bar Z)^{-1}& -(1-Z\bar Z)^{-1} Z\cr
                       \bar Z(1-Z\bar Z)^{-1} & -\bar Z(1-Z\bar Z)^{-1} Z}
.\eeq
Using the defining properties of $Z$ we can check that 
\beq
     \Phi(Z)^2=1 \quad \Phi(Z)^T=\omega^{-1} \Phi(Z) \omega \quad [\epsilon, \Phi(Z)]\in 
{\cal I}_2
,\eeq
where we used the explicit standard matrix form  of $\omega$. 
Note that these are the same conditions on our physical variable $\Phi$.
We claim that all such $\Phi(Z)$ lie on the orbit of 
$\epsilon=\Phi(Z=0)$. This is easy to see using 
\beq 
     \Phi(Z)=-g(Z)\epsilon \omega^{-1}   g(Z)^T \omega
,\eeq
which also verifies the above conditions once more.
One can see using the above identification that the action of the group on
$Z$ becomes quite simple in terms of $\Phi$,
\beq
      g\circ  Z \mapsto g\Phi g^{-1}
.\eeq
We can check  that this action preserves all the conditions on 
$\Phi$.

The manifold we have found is actually symplectic. We may define 
a natural two form,
\beq
        \Omega={i\over 4} \Tr \Phi d\Phi\wedge d\Phi
.\eeq
This formal expression should be understood as follows,
we look at vector fields at a point $\Phi$ , any such thing can be 
expressed in terms of the Lie algebra elements,
$V_u(\Phi)=[u,\Phi]$, where $u$ is an element of the Lie algebra.
then the two form becomes,
\beq
     \Omega(V_u,V_v)={i\over 8} \Tr \Phi[[u,\Phi], [v,\Phi]]={i\over 8}
\Tr \epsilon [[\epsilon, g^{-1}ug],[\epsilon, g^{-1}vg]].
\eeq
The above form shows that the trace is well-defined due to the 
Hilbert-Schmidt conditions\cite{2dqhd,rajtur,istlect}.
From  this point of view it is easy to see that  the above form is 
homogeneous, and it is closed(see \cite{2dqhd,istlect}).
Nondegeneracy can be proved at $\Phi=\epsilon$ and using homogeneity this is 
true over  the manifold.
If we look at the symplectic form at $\epsilon$ by using the 
$Z$ coordinates, we get 
\beq
        \Omega|_\epsilon=2i\Tr d\bar Z\wedge dZ
.\eeq
A short computation reveals that 
when we write $g^{-1}ug=i\pmatrix{F_1&G_1\cr -\bar G_1 &-\bar F_1}$ and same  for 
$v$, $g^{-1}vg=i\pmatrix{F_2&G_2\cr -\bar G_2 &-\bar F_2}$ we get 
\beq
     \Omega(V_u,V_v)=-{i\over 2}\Tr(G_1\bar G_2-G_2\bar G_1)
=-i{\Im}{\rm m}\Tr G_1\bar G_2
.\eeq
In fact the previous Poisson brackets come from this sympectic form, as can be checked.
We will leave the details to the reader.
We note an important point about $\Phi$, the reader can verify that 
\beq
   \Phi-\epsilon=\pmatrix{ 2Z(1-\bar Z Z)^{-1}\bar Z& -(1-Z\bar Z)^{-1}Z\cr
2\bar Z(1-Z\bar Z)^{-1} & -2 \bar Z(1-Z\bar Z)Z}\in \pmatrix{{\cal I}_1&{\cal I}_2\cr
{\cal I}_2&{\cal I}_1}
,\eeq
where ${\cal I}_1$ denotes the ideal of trace class 
operators, hence a conditional trace 
for the variable $\Phi-\epsilon$ exists.
We may therefore find moment maps which generate the underlying symmetry of the theory.
We write down the answer but do not spend much time on it since we will not make use of these
maps:
$F_u=-{1\over 2} \Tr_\epsilon u(\Phi-\epsilon)$,
here $\Tr_\epsilon A={1\over 2} \Tr(A+\epsilon A \epsilon)$.
These provide a Poisson realization of the Lie algebra.

There could be   baryonic states  in the finite $N_c$  theory given by,
\beq 
{1\over Z } \epsilon_{\alpha_1\alpha_2...\alpha_{N_c}}a^\dag_{\alpha_1}(p_1)a^\dag_{\alpha_2}
(p_2)...a^\dag_{\alpha_{N_c}}(p_{N_c})|0>
,\eeq 
where all the momenta are positive
(see the previous section).
We can measure this  baryonic  content by the half-trace of the operator 
$K$. 
We iterate again that {\it this is not a conserved quantity},  hence there is
no baryon in the  usual sense or a baryon number.
The full trace gives zero since there is no
anti-baryon. Let us see this by looking at the operator $\Phi$.
If we evaluate the trace 
$\Tr_\epsilon(\Phi-\epsilon)=2(\Tr bb^\dag -\Tr \bar b b^T)$, where
we used the appropriate group element 
$g=\pmatrix{a&b\cr \bar b &\bar a}$ to write $\Phi$.
One can see that $\Tr \bar b b^T=\overline{\Tr bb^\dag}=\Tr bb^\dag$ since $bb^\dag$
 is poisitive hermitian.
This shows that the trace is zero. In fact physically the correct one 
to take is half of this trace
as we have seen in the previous section,
so we define 
\beq
B={1\over 2} \Tr [({1+\epsilon\over 2})(\Phi-\epsilon)].
\eeq
We see that this is a positive number, which in the large-$N_c$ limit 
corresponds to the some type of baryonic  content.
The authors are unable to find a reason for this to be an integer, unlike
the 
 case discussed by Rajeev in \cite{2dqhd}, where the trace is related to
the 
 Fredholm index of the operators, thus is automatically an integer.
We face another puzzle, not only the baryonic  content is non-integer,   it is
always non-zero, that 
 is when there are mesons there are also baryons! The
limit we use seems to suggest that 
 the baryon content and mesonic states
start to mix up, since the above trace is zero only for 
 the vacuum
$\epsilon$. We are unable to resolve this issue at the moment.
Another perspective on   baryons is to think of 
the solitonic excitations of 
the gauge theory, and in our case  a nonzero trace perhaps implies 
these type of excitations. The reader may question then the validity of the 
linear approximation, since we claim that the baryon number is always 
nonzero. In the linear approximation the above trace should be taken 
zero, since it corresponds to a quadratic.

\section{Acknowledgements} 
Apart from the geometry of the phase space, this paper constitutes the first author's 
master thesis. We would like to thank Bogazici Physics Department for all the support
they have provided.
Many ideas in this paper evolved through the second author's discussions with 
S. G. Rajeev. O. T. Turgut would like to thank S. G. Rajeev for being a constant source of 
inspiration. We also gratefully acknowledge discussions with  M. Arik, 
\"O. F. Dayi, J. Gracia-Bondia, E. Langmann, J. Mickelsson,  Y. Nutku, 
C. Saclioglu and W. Zakrzewski.

\end{document}